\documentclass[10pt,conference]{llncs}
\usepackage{graphicx} % for pdf, bitmapped graphics files
\usepackage[table]{xcolor}
\usepackage{upquote}
\usepackage{multirow}
\usepackage{arydshln}
\usepackage{pgfplots}
\usepackage{tabularx} % for 'tabularx' environment

\usepackage{subfig}
  \usepackage{epstopdf}
  \DeclareGraphicsExtensions{.png,.pdf}

\graphicspath{ {images/} }
\begin{document}
\title{\LARGE \bf
Revisiting Hyper-Parameter Tuning for Search-based Test Data Generation}

\author{Shayan Zamani \and Hadi Hemmati}
\institute{Department of Electrical and Computer Engineering,
        University of Calgary, Calgary, Canada\\
        \email{\{shayan.zamani1,hadi.hemmati\}@ucalgary.ca}}
\maketitle 
\begin{abstract}
Search-based software testing (SBST) has been studied a lot in the literature, lately. Since, in theory, the performance of meta-heuristic search methods are highly dependent on their parameters, there is a need to study SBST tuning. In this study, we partially replicate a previous paper on SBST tool tuning and revisit some of the claims of that paper. In particular, unlike the previous work, our results show that the tuning impact is very limited to only a small portion of the classes in a project. We also argue the choice of evaluation metric in the previous paper and show that even for the impacted classes by tuning, the practical difference between the best and an average configuration is minor.  Finally, we will exhaustively explore the search space of hyper-parameters and show that half of the studied configurations perform the same or better than the baseline paper's default configuration.

\end{abstract}

\section{Introduction}

Since the early days of search-based software engineering (SBSE), the topic of search-based software testing (SBST) has been continuously studied and improved \cite{past-present-future}. There are now several great publicly available SBST tools such as EvoSuite for unit testing Java programs that are continuously being maintained and improved \cite{evosuite}. SBST has even gone far beyond academia and started to be deployed in large scale, e.g., in Facebook \cite{mao2016sapienz}.

In general, SBST techniques reformulate a test generation problem, e.g., maximizing branch coverage of unit tests, to an objective function and employ a meta-heuristic search technique to optimize the objective.  Examples of these meta-heuristic search techniques are hill climbing, simulated annealing, and evolutionary algorithms \cite{harmancurrent}. Evolutionary Algorithms, like Genetic Algorithm (GA), are among the most common techniques that have been used in SBST, so far. However, there are debates on whether using evolutionary algorithms or keep generation test data with random methods \cite{review4}. 

A GA starts with a set of initial population to search through the search space. Then it evolves the candidate solutions by permuting the encoded solutions with genetic and natural selection operations \cite{past-present-future}, in several iterations, until it finds the optimal solution or exhausts the search budget. Therefore, the choice of the objective function, the chromosome encoding format, and GA's input parameters (SBST's hyper-parameters) such as population size, crossover rate, mutation rate, etc. can have a significant impact on the effectiveness of the SBST technique \cite{GAbyGA}. For instance, it has been illustrated in previous work that the coverage of EvoSuite for a class varies with the values of GA's hyper-parameters \cite{Arcuri2013}.

Therefore, finding an optimal configuration of hyper-parameters could potentially, improve the SBST's effectiveness, significantly. In general, there are many parameters to be tuned for a GA. For example, Grefenstette used six parameters to tune its GA, namely, Population Size, Crossover Rate, Mutation Rate, Generation Gap, Scaling Window and Selection Function \cite{controlGA}. However, in another paper, 19 different operators or parameters are listed that contribute in the performance of a given GA and it is suggested that one should take into account the permutation of all these 19 parameters' values\cite{GAbyGA}.

Although the tuning problem is being studied in other areas frequently \cite{review1,review2,review3,Feldt:2000}, there are not many successful reports of tuning techniques in SBST. Arcuri et al. tried to find a tuned setting for EvoSuite that works better than its default for a collection of classes (SF100). However, the resulting branch coverage after tuning was less than the default configuration's results \cite{Arcuri2013}. Therefore, it is quite important for researchers in the field, as as well as practitioner, to know whether tuning is needed before using a SBST technique for test data generation and if so how much improvement a tuning method could potentially bring.

 In this paper, we partially replicate Arcuri and Fraser's paper titled ``Parameter tuning or default values? An empirical investigation in search-based software engineering'', which was published in Empirical Software Engineering journal in 2013 \cite{Arcuri2013}. The paper is one of the very few studies on the hyper-parameter tuning of SBST techniques. It includes three case studies, where the first one focuses on illustrating the impact of tuning. The two other case studies then investigate the effectiveness of a proposed tuning method (which showed no improvement compared to default settings). In our study, we only focus on the impact of tuning, thus only the first case study of their paper (from now on called ``the baseline paper'') will be (partially) replicated.
 
In particular, we focus on the first two research questions of the first case study, in the baseline paper, where the main findings are:
\begin{itemize}
\item  ``Different parameter settings cause very large variance in the performance'' 
\item ``Default parameter settings perform relatively well, but are far from optimal''.
\end{itemize}

Here we argue that the conclusions are taken from the set of 20 handed picked classes, which do not represent the entire project's classes. We exhaustively study the impact of tuning, with the similar hyper-parameter search space as the baseline paper (1,200 different configurations), on 117 classes of three random projects from the very same SF100 and show that different parameter settings do NOT have the significant impact that is claimed. We show this first by looking into all classes vs. a handpicked set and second by measuring improvements using the raw coverage (the number of extra branches covered when using a different configuration) rather than a relative measure used in the baseline paper. Together we show that the impact of tuning is much less than the reported ones. In addition, we show that the true impact is limited to the maximum improvement ranges (called ``potentials'' in this paper) in the project-level, which are 64\% of class-level potentials.  Finally, we analyze the distribution of the entire 1,200 configurations of hyper-parameters in terms of their effectiveness and show that half of the configurations are performing at least as well as the default configuration. This means that even a randomly selected configuration would have a 50\% chance to be better than the default setting, but the issue is the improvements are minor (maximum 12 extra branches per project), anyways. 

The main contributions and findings of this paper are as follows:
\begin{itemize}
\item Running an exhaustive search with over 2 million configuration/class pair evaluations, to study the impact of hyper-parameter tuning.
\item Replicating the previous study on tuning and showing (clarifying) that, in contrary to the reported results, on average, different parameter settings cause no change in coverage at all on most (81\%) classes.
\item Showing that, in some cases, a relative coverage measure may not be the best metric to explain the potentials of tuning. Tuning for only 12 extra branches per project would be reported as 52\% average improvement per class using the relative coverage.
\item Showing that half of the possible configurations perform as well or better than the default configuration, but overall, the practical improvements are insignificant in most cases. 
\end{itemize}

\section{Empirical Study}
In this section, we will explain the details of our experiment design and results.

\subsection{Objective and Research Questions} \label{rq}

Our objective is to revisit the previous study on the impact of tuning on SBST, by partially replicating our baseline paper, introduced in the introduction section. Our main hypotheses are that a) not all classes in a project are significantly sensitive to parameter tuning, b) the improvement of coverage is magnified in previous studies, and c) most configurations are already good and don't leave much room for improvement in the SBST context. To investigate the above hypotheses, we design the following research questions: 

\begin{itemize}
\item {\bf RQ1:} {\it What portion of classes in a project would be sensitive to hyper-parameter tuning?}\\
The idea of this question is to first identify classes that won't be affected at all no matter what configuration will be used.
\item {\bf RQ2:} {\it To what extent code coverage of classes within a project may change, when the hyper-parameters of SBST techniques change?}\\
Knowing the answer to RQ1, we now need to know how much potential improvement one can gain by tuning, to decide whether tuning is even worthwhile (if the portion of sensitive classes is small the potentials are negligible, tuning is not justifiable). 
\item {\bf RQ3:} {\it How are different hyper-parameter configurations (including the default from the baseline paper) compared in terms of their resulting code coverage?}\\
The goal of this RQ is to dig deeper into the effectiveness of different configurations and see where a default configuration sits comparing to a median setting. 

\end{itemize}

\subsection{Experiment Design}

\subsubsection{Subjects under study:} \label{subject}
We have selected three random projects from the SF100 Java benchmark (which is a well-known dataset in SBST and have been used in the previous work. It is also the same source for our baseline paper \cite{Arcuri2013}), namely, JSecurity, Geo-Google, and JOpenChart. We made sure that the sizes of our selected projects are around or greater than the median project size within SF100, which is 35 classes\cite{SF100}, per project, so that we don't study only the trivial projects, by chance. In addition, we checked that the average number of branches per class in our selected projects are around the median value of SF100 projects, which is 18 branches (See table \ref{sf100-info2}).
Therefore, by considering these two measurements, we believe that our random selected classes are representative of other SF100 projects.

The summary of SF100 projects' properties is available in the table \ref{sf100-info} from the information available in \cite{SF100}:

\begin{table}[h!]
\centering
\begin{tabular*}{\textwidth}{||c@{\extracolsep{\fill}}c@{\extracolsep{\fill}}c@{\extracolsep{\fill}}c@{\extracolsep{\fill}}c||} 
 \hline
   & Min & Median & Average & Max \\
 \hline\hline
 \# of Classes per Project & 1 & 35 & 87.84 & 2189\\
 \# of Branches per Class & 0 & 18 & 33.20 & 2480\\
\hline
\end{tabular*}
\caption{Summary of SF100 projects statistics}
\label{sf100-info}
\vspace{-15mm}
\end{table}

\begin{table}[h!]
\centering
\begin{tabular*}{\textwidth}{||c@{\extracolsep{\fill}}c@{\extracolsep{\fill}}c@{\extracolsep{\fill}}c||} 
 \hline
  Project & \# of Classes & \# of Branches & Average \# of Branches per Class \\
 \hline\hline
 JSecurity & 72 & 998 & 13.86\\
 Geo-Google & 52 & 1344 & 25.84\\
 JOpenChart & 38 & 693 & 18.24\\
\hline
\end{tabular*}
\caption{Statistics of randomly selected projects from data in \cite{SF100}}
\label{sf100-info2}
\vspace{-15mm}
\end{table}

\subsubsection{SBST tool:} \label{tool}
In order to evaluate the tuning techniques, we use the well-known open source SBST tool, EvoSuite \cite{evosuite}, \cite{test-suite}. EvoSuite is also the tool that was used in the baseline paper \cite{Arcuri2013}.  EvoSuite accepts Java bytecode of a class and creates a test suite that maximizes different criteria (e.g., branch coverage) using a Genetic Algorithm for optimization \cite{benchmark}.
The GA parameters it employs are configurable. If no specific parameter is passed to EvoSuite, it will use a default setup that we refer to as ``EvoSuite defaults'' or the ``baseline paper defaults''. EvoSuite defaults have been selected by following guidelines, best practices, and experimentation, and have shown to be quite good and reasonable in the previous study \cite{Arcuri2013}.

\subsubsection{Measurements:} \label{measures}
In this study, we use code (branch) coverage as our test adequacy measure, to be consistent with our baseline paper \cite{Arcuri2013}.
In addition to the raw coverage data (number of the covered branches as well as their percentages), we also report ''Relative Coverage``, which is suggested \cite{statistics} and used in the baseline paper \cite{Arcuri2013}. The rationale of using the relative coverage in the baseline paper was that ``using the raw coverage values for parameter setting comparisons would be too noisy. Most branches are always covered regardless of the chosen parameter setting, while many others are simply infeasible''.

Therefore, for a given configuration with resulting branch coverage equal to $b$ on class $c$, we report Relative Coverage rc(b,c) as defined below:
\begin{displaymath}
Relative Coverage = \frac{b - min}{max - min}
\end{displaymath}
where b is the number of covered branches, and min and max are the minimum and maximum number of branches covered in the class $c$ over all experimented configurations.

During the experiments, we report both raw coverage and the relative coverage and discuss this choice of metric. 

Among all classes within a project, there are some classes that have the same coverage for any configuration at any iteration, and their $max$ and $min$ values are the same. We call them as \textbf{insensitive classes}. 

\subsubsection{Experiment Procedure:} \label{experiment}
The experiments investigate the sensitivity of the three projects under study, in terms of branch coverage, when the GA hyper-parameters changes. Basically, we define a set of limited values per GA parameter (in EvoSuite) and run an exhaustive search over this search space. We then run EvoSuite per configuration and calculate branch coverage for all classes within each project.

The search space consists of the combination of 5 most important parameters of GA in EvoSuite, i.e. population size, crossover rate, elitism rate, selection function, and parent replacement check. In the tuning literature, there are some cases in which more parameters are considered to be tuned \cite{GAbyGA,controlGA}. However, due to the extreme cost of exhaustive search and to be consistent with our baseline paper, we limit the tuning to these five parameters. 

 In the following, there is a brief explanation of the parameters of our interest to be tuned:
\begin{enumerate}
    \item Crossover Rate: It is the probability with which two candidates selected from the parent generation are crossed over. 
    \item Population Size: This parameter indicates how many individuals exist in each generation and due to mutation and crossover operations the population remains constant while evolving.
    \item Elitism Rate: This parameter determines how many or what percentage of top individuals are exempted from any crossover or mutation during evolution and are directly passed to the next generation without any modification. 
    \item Selection Function: This parameter specifies the mechanism with which individuals of a population are selected for the purpose of reproduction operations. In oppose to the other mentioned parameters, this one is not numerical and is a nominal variable. Three types of known selection methods are roulette wheel selection, tournament selection, and rank selection.\\ 
    In the roulette wheel selection method, individuals with more fitness score are more probable to be selected.\\
    In tournament selection, based on the tournament size a number of individuals are selected uniformly and this method does not weight the selection probability regarding fitness score.\\
    Rank selection considers fitness score into its selection method; however, unlike roulette wheel selection, the probability is not proportional to fitness score, rather it is based on the rank of individuals.\\
    Therefore, the fittest individuals do not dominate the selection like what happens in the tournament approach.
    \item Parent Replacement Check: If this parameter is considered in the genetic algorithm, it checks the two off-springs, which are generated in the reproduction phase, against their parents. If they do not show an improvement in fitness score compared to at least one of their parents, they are not included in the next generation, and the algorithm continues with the parents in the next generation \cite{Arcuri2013}.
\end{enumerate}

Following the baseline paper (the settings from its first case study) \cite{Arcuri2013}, we also have limited the values per parameter to a small discretized sub-samples, as follows:
\begin{itemize}
    \item Crossover rate: {0, 0.2, 0.5, 0.75, 0.8, 1} (6 cases)
    \item Population size: {4, 10, 50, 100, 200} (5 cases)
    \item Elitism rate: {0, 1, 10\%, 50\%} (4 cases)
    \item Selection: roulette wheel, tournament with size either 2 or 10, and rank selection with bias either 1.2 or 1.7 (5 cases)
    \item Parent replacement check: activated or not (2 cases)
\end{itemize}

Thus, the search space under exploration in this study includes $6 \times 5 \times 4 \times 5 \times 2 = 1,200$ different combinations of GA parameters, used by EvoSuite.

During an experiment, we take a pair of one configuration and one class, each time, and evaluate this pair's coverage using EvoSuite, with a search budget of two minutes, which is a realistic amount in practice and was used in the baseline paper as well. In order to address randomness, we repeat each evaluation 10 times. The average coverage of these repetitions is then reported as the pair's coverage.

There are 177 classes in total, in the three projects under study. Considering 10 repetitions of evaluating these classes on a search space of 1,200 configurations, where each evaluation takes two minutes, this analysis would take $177 \times 10 \times 1200 \times 2$ minutes equals to more than 8 years, on a single core machine. Therefore, we used computer clusters to run this amount of computation in parallel.
The clusters in use were ComputeCanada (Graham with 32-core instances and Cedar with 48-core instances) and Cybera (8-core instances), which are all Linux-based systems, summed up to 360 nodes.

Note that all the scripts and output results are publicly available  \footnote{https://github.com/sea-lab/EvoSuiteTuning}.

\subsection{Results} \label{results}
In this section, we will report and explain the results of the experiments per research question.
\subsubsection{RQ1 (Insensitive classes):}
As discussed in Section \ref{measures}, a large proportion of classes within a project may be insensitive to hyper-parameter tuning, which affects the usefulness of tuning in practice. In RQ1, we report this proportion per project in our study.
Looking at Table \ref{table:1} we see that 86\% (73 out of 85), 95\% (53 out of 56), and 50\%(18 out of 36) of classes were insensitive to hyper-parameter tuning in our three projects. 
 The number of classes that we observed in the projects are different from what is presented in table \ref{sf100-info2} which were taken from \cite{SF100}. This is due to the reason that the projects are still changing, and versions are different.
 \begin{table}[h!]
\centering
\begin{tabular*}{\textwidth}{||c@{\extracolsep{\fill}}c@{\extracolsep{\fill}}c@{\extracolsep{\fill}}c||} 
 \hline
  Project & \#Classes & \#Insensitive Classes & Proportions \\
 \hline\hline
 JSecurity & 85 & 73 & 0.86\\
 Geo-Google & 56 & 53 & 0.95\\
 JOpenChart & 36 & 18 & 0.50\\
 \hline
 Total & 177 & 144 & 0.81\\
\hline
\end{tabular*}
\caption{The proportion of insensitive classes per projects.}
\label{table:1}
\vspace{-10mm}
\end{table}
 The high number of insensitive classes in each project highlights that although tuning SBST techniques for some classes may be useful (see RQ2), the coverage of most classes in a given project will not be affected by tuning. Thus, applying an umbrella tuning on all classes of a project may not be effective, and the impact of tuning may be only limited to a small portion of the project.
 
 This is in contrary with this generic claim from the baseline paper \cite{Arcuri2013}: ``Different parameter settings cause very large variance in the performance''. The issue with that claim is that it is based on the 20 manually selected classes, where the tuning was indeed effective. The justification for this selection is given as ``We, therefore, selected classes where EvoSuite used up its entire search budget without achieving 100\% branch coverage, but still achieved more than 80\% coverage.'' Though this might be a fine selection criterion to detect the sensitive classes, the problem is that the conclusions are generic and do not consider the significant number of insensitive classes.
 
\begin{figure}[ht!]%
\vspace{-7mm}
    \centering
\noindent\makebox[\textwidth]{%
    \subfloat[JSecurity]{
    \begin{tikzpicture}
\begin{axis}[
    ymin=0, ymax=71,
    minor y tick num = 3,
    height=4cm,
    width=4.6cm,
    area style,
    xlabel=Coverage,
    ylabel=Class Counts,
    ]
\addplot+[ybar interval,mark=no] plot coordinates { (0,4)(0.2,7)(0.9,55)(1,0)};
\end{axis}
\end{tikzpicture}
    }%
    \subfloat[Geo-Google]{
    \begin{tikzpicture}
\begin{axis}[
    ymin=0, ymax=53,
    minor y tick num = 3,
    height=4cm,
    width=4.6cm,
    area style,
    xlabel=Coverage,
    % ylabel=Class Counts,
    ]
\addplot+[ybar interval,mark=no] plot coordinates { (0,1)(0.2,1)(0.9,51)(1,0)};
\end{axis}
\end{tikzpicture}
}%

    \subfloat[JOpenChart]{
    \begin{tikzpicture}
\begin{axis}[
    ymin=0, ymax=18,
    minor y tick num = 3,
    area style,
    height=4cm,
    width=4.6cm,
    xlabel=Coverage,
    % ylabel=Class Counts,
    ]
\addplot+[ybar interval,mark=no] plot coordinates { (0,1)(0.2,0)(0.9,17)(1,0)};
\end{axis}
\end{tikzpicture}
    }}%
    \caption{Distribution of 144 insensitive classes in 3 projects, over the coverage range.}%
    \label{insensitives}%
    \vspace{-5mm}
\end{figure}
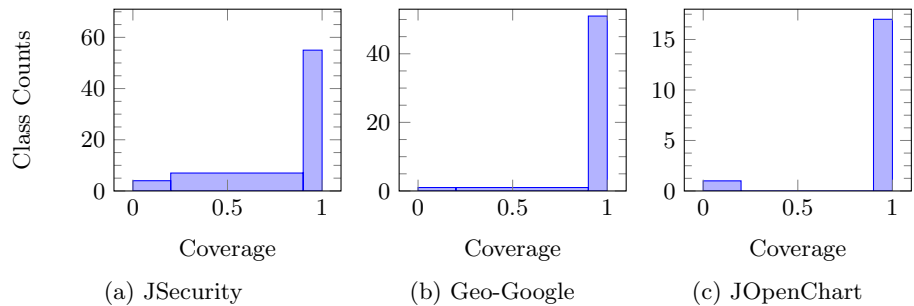

In fact, our study on 177 classes shows that a blanket tuning over all classes of a project will NOT have a very large variance in the SBST technique's performance. Insensitive classes are mainly the ones that are trivial and easy for the SBST tool to evaluate. The analysis of insensitive classes in the projects under study shows that 85\% (123 out of 144) of them have branch coverage more than 0.9 while 4\% (6 out of 144) of them are very difficult to cover and their branch coverage is lower than 0.2, regardless of the configuration. Figure \ref{insensitives} summarizes the distribution of classes per project, over the coverage range.

So our conclusion is that tuning will NOT make a big difference in the coverage of the entire project. So, in the SBST context, the tuning effort should be focused only on the sensitive classes. Obviously, this first requires a systematic approach to detect such classes, before tuning, and second a metric that only considers those classes when evaluating the improvements, which we discuss more in RQ2.    
 
 \begin{center}
\fbox{\parbox{4.75in}{\centering
On average, more than 81\% (144 out of 177) of classes in three projects under study were insensitive to 1,200 different configurations of GA hyper-parameters.}}
\end{center}

\subsubsection{RQ2 (Evaluation metric):}
As discussed in the RQ1 results, the coverage of most classes in a project are indifferent to the hyper-parameter configuration of the SBST technique. In this section, we will assess the potential of the remaining sensitive classes to improve their branch coverage by SBST tuning, with two metrics: number of covered branches and relative coverage. Then, we will discuss how useful are these metrics, in this context. 
\begin{table}[h!]
\vspace{-5mm}
\centering
\resizebox{\textwidth}{!}{\begin{tabular}{clc|ccc|ccc}
 \hline
 & &  & \multicolumn{3}{c}{Covered Branches} & \multicolumn{3}{|c}{Relative Coverage}\\\cline{4-9}
Project & Class Names & Branch & Median & Worst & Best & Median & Worst & Best\\
\hline
\multirow{15}{*}{JSecurity}
& DefaultWebSessionFactory            & 75 & 55.6 & 50.40 & 58    & 0.71 & 0.54 & 0.78 \\
& AbstractSessionManager              & 89 & 65.4 & 58.67 & 70.8  & 0.48 & 0.24 & 0.67 \\
& DefaultSessionManager               & 33 & 21   & 16.50 & 25.4  & 0.52 & 0.33 & 0.71 \\
& MemoryAuthenticationDAO             & 24 & 20.75 & 9.50  & 24    & 0.81 & 0.15 & 1.00 \\
& AbstractAuthenticator               & 47 & 28.4 & 21.60 & 29    & 0.96 & 0.56 & 1.00 \\
& SimpleAuthorizationContext          & 57 & 52 & 51.25 & 53    & 0.38 & 0.28 & 0.50 \\
& ThreadContext                       & 24 & 17 & 16.67 & 18.25 & 0.17 & 0.11 & 0.38 \\
& DAOAuthenticationModule             & 12 & 5  & 5.00  & 10    & 0 & 0.00 & 1.00 \\
& DefaultSessionFactory               & 9  & 5  & 5.00  & 6.2   & 0 & 0.00 & 0.30 \\
& MemorySessionDAO                    & 24 & 22.8 & 21.75 & 24    & 0.60 & 0.25 & 1.00 \\
& DelegatingSession                   & 20 & 17 & 17.00 & 17.6  & 0 & 0.00 & 0.20 \\
& ModularAuthenticator                & 18 & 8  & 8.00  & 8.6   & 0 & 0.00 & 0.20 \\
& SimpleSessionEventSender            & 18 & 18 & 17.60 & 18    & 1.00 & 0.80 & 1.00 \\
& ActiveDirectoryAuthenticationModule & 14 & 7  & 6.50  & 7     & 1.00 & 0.75 & 1.00 \\
& SimpleAuthenticationEventSender     & 29 & 28 & 27.20 & 28    & 1.00 & 0.60 & 1.00 \\
& WebUtils                            & 15 & 15 & 14.75 & 15    & 1.00 & 0.75 & 1.00 \\
& LdapAuthenticationModule            & 27 & 16 & 15.75 & 16    & 1.00 & 0.75 & 1.00 \\
& AnnotationAuthorizationModule       & 13 & 8.6  & 8.00  & 9     & 0.60 & 0.00 & 1.00\\
\hdashline
Average & & 30.44 & 22.81 & 20.62 & 24.32 & 0.57 & 0.34 & 0.76\\
\hline
\multirow{3}{*}{Geo-Google}
& AddressToUsAddressFunctor & 30 & 22 & 8.5  & 22   & 1.00 & 0.21 & 1.00 \\
& GeoAddressStandardizer    & 38 & 27 & 23.6 & 28.5 & 0.87 & 0.64 & 0.97 \\
& MappingUtils              & 8  & 6.1  & 5.4  & 6.6  & 0.37 & 0.13 & 0.53\\
\hdashline
Average & & 25.33 & 18.37 & 12.5 & 19.03 & 0.74 & 0.33 & 0.83\\
\hline
\multirow{18}{*}{JOpenChart}
& CoordSystemUtilities             & 92 & 70.4 & 39.8  & 89.33 & 0.69 & 0.25 & 0.96 \\
& RadarChartRenderer               & 22 & 2  & 2       & 6.00  & 0.00 & 0.00 & 0.20 \\
& CoordSystem                      & 71 & 65.2 & 59.6  & 69.00 & 0.75 & 0.37 & 1.00 \\
& DefaultChart                     & 20 & 6  & 5.8     & 15.80 & 0.07 & 0.05 & 0.72 \\
& BarChartRenderer                 & 16 & 7.6  & 4     & 11.80 & 0.40 & 0.14 & 0.70 \\
& InterpolationChartRenderer       & 17 & 12 & 8.75    & 13.40 & 0.67 & 0.40 & 0.78 \\
& AbstractChartDataModel           & 37 & 27.8 & 22.75 & 31.40 & 0.62 & 0.16 & 0.95 \\
& DefaultChartDataModelConstraints & 30 & 28.8 & 22.2  & 30.00 & 0.89 & 0.29 & 1.00 \\
& StackedChartDataModelConstraints & 54 & 51 & 46.8    & 51.00 & 1.00 & 0.58 & 1.00 \\
& LineChartRenderer                & 17 & 12 & 9.2     & 13.40 & 0.60 & 0.32 & 0.74 \\
& PieChartRenderer                 & 14 & 2  & 2       & 5.60  & 0.00 & 0.00 & 0.36 \\
& StackedBarChartRenderer          & 18 & 13.6 & 7.8   & 15.80 & 0.66 & 0.08 & 0.88 \\
& PlotChartRenderer                & 11 & 9.6  & 7     & 10.00 & 0.93 & 0.50 & 1.00 \\
& DefaultChartDataModel            & 34 & 34 & 32.6    & 34.00 & 1.00 & 0.72 & 1.00 \\
& AbstractRenderer                 & 7  & 6.2  & 3.5   & 7.00  & 0.84 & 0.30 & 1.00 \\
& ChartEncoder                     & 9  & 4.8  & 3     & 6.00  & 0.60 & 0.00 & 1.00 \\
& Legend                           & 12 & 11.6 & 10.25 & 12.00 & 0.87 & 0.42 & 1.00 \\
& AbstractChartRenderer            & 16 & 15.00 & 14.8 & 15.00 & 1.00 & 0.80 & 1.00\\
\hdashline
Average & & 27.61 & 21.08 & 16.77 & 24.25 & 0.64 & 0.30 & 0.85\\
\hline 
Overall & & 31.14 & 23.48 & 19.73 & 25.87 & 0.67 & 0.34 & 0.86\\
\hline
\end{tabular}}
\caption{Sensitive classes in the projects under study, and their coverage in terms of the number of covered branches and the relative coverage.}
\label{RQ2}
\vspace{-10mm}
\end{table}

Table \ref{RQ2} reports the number of total branches per class (the Branch column) and summarizes the branch coverage as the number of covered branches vs. relative coverage (explained in Section \ref{measures}). For each category, it reports the median, the worst, and the best numbers overall 1,200 $ \times $ 10 configuration evaluations, per class. The last row of each project summarizes all columns per project. 

Following the relative metric, suggested in the baseline paper, we can conclude that the range of relative coverage per class is huge when looking at the best vs worst relative coverage (On average 42\%, 50\%, and 55\% in JSecurity, Geo-Google, and JOpenChart projects, respectively). However, if we look at the raw coverage numbers the range between the best and the worst configuration, with respect to the number of covered branches are pretty small (3.7, 6.53, and 7.48 in JSecurity, Geo-Google, and JOpenChart projects, respectively). These numbers can be minimal for some classes, e.g., in the JSecurity project, there are 8 classes (out of 15 classes) where the difference between the best and the worst configurations is less than one branch (which practically can be called an insensitive class).

In other words, although the relative coverage metric shows a great potential (52\%) for improvement using a tuning technique, the actual raw numbers reveal that the practical impact is limited to a few branches (on average 6.14 extra ). This is equal to (6.14/31.14) 19.7\% improvement on raw branch coverage. Although even the 6.14 extra branches might be among buggy ones and thus a good tuning would in fact result in extra bug detection, but the point we make here is that the measurement should be reflective of the real-world effect. If the raw potential is 19.7\% (regardless of how many more bugs potentially can be detected by such a tuning) we should not say the potential is 52\%. This artificially exaggerates expectations from a tuning method.

Another point is that the impact of tuning is not going to be on the scale of the range of coverage as reported above (The Best - The Worst). In practice, choosing the worst configuration is rare.  Even a random configuration would be better than the median results in 50\% of the times. Thus, a more reasonable comparison is to set the expectations for improvement between the best and the median, not the worst. Following this approach, the improvement in raw branch coverage per class would be even smaller (25.87 - 23.48 = 2.39 branches, equals to 2.39/31.14 = 7.7\% potential coverage improvement for each class).

Therefore, we can conclude that although using relative coverage can avoid noises in the results, in some cases, it exaggerates the effectiveness of SBST tool tuning while there are only a few extra branches to be covered.

Following the above discussion, in RQ3, we will look deeper into the distribution of configurations and their corresponding results. 

\begin{center}

\fbox{\parbox{4.75in}{\centering
Looking at relative coverage are not always helpful in terms of measuring tuning potentials. On average, an extra 6 branches on a total of 31 branches in sensitive classes would be reported as 52\% improvement in relative coverage -- In addition, the potential improvement on raw branch coverage when comparing the best and the median is just 2.39 branches, per class. }}

\end{center}

\subsubsection{RQ3 (Distribution of configurations):}
In RQ2, we listed the best/ worst/ median configurations for each sensitive class of projects under study. When it comes to tuning an SBST tool on a project scale, the problem changes a bit. The goal is no longer finding the best configuration per class. It is rather finding one single tuned configuration that works the best over all classes of the project. Note that these two (the class-level and the project-level best configurations) are different. In many cases, It might not be possible to have a configuration that works best for all classes. 

Therefore, it is obvious that it might not be possible for a tuning method to be as good as the best configuration as reported in RQ2. So, we define a ``Maximum/Minimum'' configuration as the best/worst possible configuration in a project-level, as the configuration that results in the highest total branch coverage over all classes in the project. This follows the baseline paper's definition of ``optimal'' (equal to our ``Maximum'') configuration, as well. 

Now to see the real impact of tuning, in table \ref{RQ3-1}, we report the Maximum and Minimum results per project. We also report the Best and Worst from RQ2 aggregated on a project-level to show that these two measures are different.

Overall, we can observe that the range of feasible coverage (Maximum - Minimum ) of a given project is quite smaller than the range of potential coverage reported in RQ2. Looking at table \ref{RQ3-1}, the feasible ranges of coverage for projects JSecurity, GeoGoogle, and JOpenChart are only 54, 86, and 52 (on average 64) percents of the potential ones, reported in RQ2, respectively.

Thus, when it comes to assessing the impact of tuning, we have to measure it with raw coverage measures and look at the feasible coverage range bounds.

In practice, a typical SBST tool would have a default configuration, which tuning's goal is to improve its performance. So the next question is to see how EvoSuite default configuration performs in comparison to the other configurations in the search space.

The current values of EvoSuite default configuration for the hyper-parameters of our interest are as follows:
\begin{itemize}
    \item Crossover rate: 0.75
    \item Population size: 50
    \item Elitism rate: 1
    \item Selection: rank selection with bias either 1.7
    \item Parent replacement check (activated)
\end{itemize}
\begin{table*}[h]
\vspace{-8mm}
\centering
\resizebox{\textwidth}{!}{\begin{tabular}{||c|c|c c|c c|c c||} 
 \hline
  & & \multicolumn{2}{c}{Project-level Potential Range} & \multicolumn{2}{|c|}{Class-level Potential Range} & & \\\cline{3-6}
  Project & Total & Maximum & Minimum & All Best & All Worst & Median & Default \\
 \hline\hline
 JSecurity & 1093 & 828.20 & 792.33 & 843.85 & 777.13 & 818.0 & 822.6 \\
 Geo-Google & 1408 & 1371.6 & 1354.8 & 1372.1 & 1352.5 & 1370.0 & 1370.3 \\
 JOpenChart & 795 & 667.8 & 579.55 & 697.53 & 562.85 & 644.0 & 639.7\\

\hline
\end{tabular}}
\caption{The number of branches per project, the class-level and project-level potential ranges of covered branches, and the median and default performance.}
\label{RQ3-1}
\vspace{-8mm}
\end{table*}
Looking at last two columns of Table \ref{RQ3-1}, we will see that although the default configuration of EvoSuite is working well and is very close to the Maximum (optimal) coverage of each project (only misses less than 11.66 branches, on average), the median covered branches of all configurations is also very close (missing only 11.86 branches, on average), and performs even better than default for the JOpenChart project. 

\begin{figure}[ht!]%
\vspace{-8mm}
    \centering
    % \noindent\makebox[\textwidth]{%
    \subfloat[JSecurity]{
    \begin{tikzpicture}
\begin{axis}[
    ymin=0, ymax=700,
    minor y tick num = 3,
    height=4.25cm,
    width=6cm,
    area style,
    xlabel=Number of Covered Branches,
    ylabel=Config. Counts,
    ]
\addplot+[ybar interval,mark=no] plot coordinates { (792, 13) (796, 30) (800, 29) (804, 15) (809, 36) (813, 253) (817,579) (821,219) (825,26) (828,0)};
\end{axis}
\end{tikzpicture}
    }%
    \subfloat[Geo-Google]{
    \begin{tikzpicture}
\begin{axis}[
    ymin=0, ymax=1000,
    minor y tick num = 3,
    height=4.25cm,
    width=6cm,
    area style,
    xlabel=Number of Covered Branches,
    ylabel=Config. Counts,
    ]
\addplot+[ybar interval,mark=no] plot coordinates { (1355, 5) (1357, 17) (1359, 15) (1361, 20) (1363, 21) (1365, 6) (1367,43) (1369,766) (1371,307) (1373,0)};
\end{axis}
\end{tikzpicture}
}%

    \subfloat[JOpenChart]{
    
    \begin{tikzpicture}
    
\begin{axis}[
    ymin=0, ymax=400,
    minor y tick num = 3,
    area style,
    height=4.25cm,
    width=6cm,
    xlabel=Number of Covered Branches,
    ylabel=Config. Counts,
    ]
\addplot+[ybar interval,mark=no] plot coordinates { (580, 28) (589, 22) (598, 30) (606, 14) (621, 56) (629, 174) (637,300) (645,345) (653,202) (661,29) (668,0)};
\end{axis}
\end{tikzpicture}
    }%
    \caption{Distribution of 1,200 configurations in the search space over coverage, per project.}%
    \label{configurations}%
    \vspace{-10mm}
\end{figure}
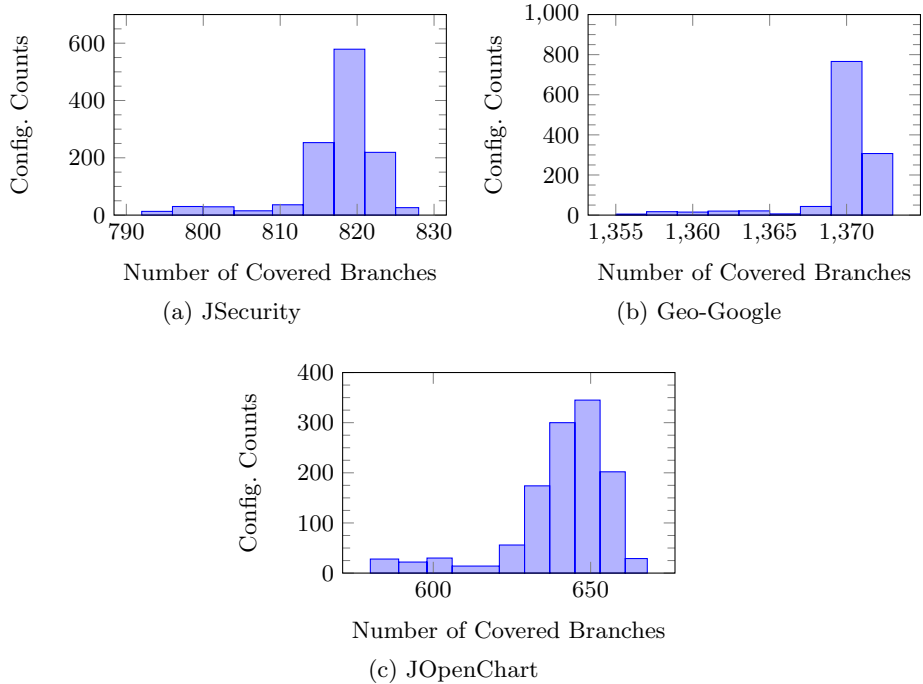

This suggests that 50\% of configurations in our 1,200-member search space, i.e. 600 configurations, are working with a performance very close to or better than the default, which means one has at least 50\% chance to select a configuration as good or better than the default, randomly, without any tuning. In figure \ref{configurations}, the entire distribution of configurations and their yielded coverage is illustrated, for all three projects.

\begin{center}
\fbox{\parbox{4.75in}{\centering
Tuning in practice is done on the entire project level, which on average has a 36\% less potential for improvement compared to what was reported in RQ2 -- As the observed median coverage suggests, half of the configurations are performing as well or better than the default configuration.}}

\end{center}

\subsection{Threats to Validity}

In terms of construct validity, the metrics that we used in our experiments are the number of covered branches and relative coverage which both are very common metrics in the context of software search-based testing to illustrate branch coverage. In RQ2, we compared the results of these two metrics and avoided the relative coverage reported in the baseline paper. 

In terms of conclusion validity, we address the randomness of the SBST coverage results by repeating each experiment 10 times with different random seeds and taking averages. Note that since we do not directly compare different techniques and rather just show the ranges a statistical significant test was not applicable.

To minimize internal validity, we replicated the baseline paper as much as possible. We used the same tool, dataset, and hyper-parameters. The only part which was slightly different was that we used the current default values of EvoSuite, whereas the baseline paper used the default values at the time. Although the overall results are not that different, we wanted to make sure that we use the best default the tool comes with when comparing it with median configuration results. 
In addition, one possible threat is changing the default configuration from what is used in \cite{Arcuri2013} to what is used in the current version of EvoSuite (i.e. 1.0.6). Our assumption is that given the wealth of projects that use EvoSuite on SF100 and beyond, over the years from the publishing time of the baseline paper, the new baseline has improved and provides even better results (or at least equal) to those reported in the baseline paper. So it is safe to focus only on the new version of EvoSuite defaults.

Regarding external validity, more empirical studies are needed to generalize the results. Our results are limited to SF100 (not even covering that all). So we need to replicate the study on more SF100 projects and beyond. In addition, we are limited to GA-based SBST (the EvoSuite implementation). Therefore, more experiments are needed where other GA or non-GA-based SBST tools are studied, with respect to the tuning impact. However, given the extreme cost of this type of study, our results are very valuable. The replication nature of the paper is also indirectly helping on the generalization of this type of studies.  

\section{Related Works}
One of the most influential works in tuning in the context of search-based software testing is the empirical study of effect of tuning evolutionary algorithms on \textit{test data generation} tools \cite{SBSEtuning}, which was extended later on in the baseline paper for our study \cite{Arcuri2013}. Their preliminary work focuses on 20 random Java classes and uses the relative coverage metric. They noted that there is a high variance of coverage when a different configuration is set for the SBST tool \cite{SBSEtuning}. Later, knowing that tuning is effective for improving the coverage, they applied response surface methodology tuning method on 10 large-scale projects with 609 classes in total to assess only 280 configurations. But, the raw coverage of the tuned configuration was found to be less than default coverage. They reported that the tuning method in use was not working in the SBST context \cite{Arcuri2013}. Later, Kotelyanskii et al. replicated this study using Sequential Parameter Optimization Toolbox (SPOT) method and confirmed that the default setting for EvoSuite is performing well, and tuning cannot outperform it \cite{replication}. In contrast, in our study, we studied 177 Java classes from three projects on the whole set of configurations rather than on a few configurations.

While the aforementioned papers are the closest ones in terms of the context of our study, there are many other papers that used and confirmed their finding about effectiveness of tuning in other applications of SBSE. For example, in the configuration of Software Product Lines problem based on stakeholder needs, two meta-heuristics were evaluated with different hyper-parameter settings. It was found that the performance of algorithms depends on the hyper-parameter settings \cite{replicate}.
Parameter tuning of machine learning approaches to solving \textit{software effort estimation} problem has been also studied, and it was shown that parameter settings make a difference in the results of machine learning performance \cite{MLtuning}.
In addition, it was found that tuning machine learning defect predictors can improve the performance, and it can even change the decisions on what are the important factors of software development \cite{defect}.
In another study on 6 \textit{clone detection} tools that are used widely, it was shown that tool configuration can improve the performance \cite{cloneDetection}.
In our study, however, we claim that this dependency to the hyper-parameter settings in the context of our interest doesn't change the results significantly and is limited to covering only a few more branches.

\section{Conclusion and Future Work}
This paper revisits the problem of hyper-parameter tuning in SBST, studied in a previous publication. Studying 177 Java classes from 3 random projects from SF100, we observed that 81\% of classes are insensitive to tuning. Moreover, the evidence from this study implies that the relative coverage improvement, used in the baseline paper, may unhelpfully exaggerate the effectiveness of tuning. Exhaustively searching through 1,200 configurations in the hyper-parameters search space, we conclude that not only EvoSuite default but also half of the configurations are covering most of the branches missing only about 12 branches per project compared to the best feasible coverage.

Regardless of the low potentials for tuning observed in this study, the next main observation is that the potentials are much higher in the individual class-levels than the entire project tuning. Thus for future works, we will try to devise a tuning method that looks at the static features of classes and tunes the configurations in class-level rather than project-level. Moreover, we will try more GA-based (e.g., EvoMaster \cite{evomaster}) and non-GA-based SBST (e.g., \cite{not_evolutionary} ) techniques to confirm and generalize our findings and will extend our study to more Java projects from SF100 and beyond. 
% \section*{Acknowledgement}
% This work is partially supported by the Natural Sciences and Engineering Research Council of Canada [RGPIN/05108-2014].

\bibliographystyle{abbrv}
\bibliography{andrea}

\end{document}